# Near-field Holographic Retrieval of an Isolated Subwavelength Hole in a Thin Metallic Film

Jun Xu[1,2], Hyungjin Ma[3], and Nicholas X. Fang[1*]

[1]Department of Mechanical Engineering, Massachusetts Institute of Technology, Cambridge, MA, 02139
[2]Department of Mechanical Science and Engineering, University of Illinois at Urbana-Champaign, Urbana, IL, 61801
[3]Department of Physics, University of Illinois at Urbana-Champaign, Urbana, IL, 61801

Using a holographic approach, we experimentally study the near-field intensity distribution of light squeezed through an isolated subwavelength plasmonic hole in a thin metallic film. Our experiments revealed an in-plane electric dipole moment excited near the isolated hole. By analyzing the fringe patterns formed between the in-plane dipole and plane wave illumination, both the transmission coefficient and phase shift of the dipole can be retrieved. We also observed opposite phases of the excited dipoles from the subwavelength dent and protrusion in the metallic film, in good agreement with the prediction from our model. Our approach can be used to study the microscopic process of the light-structure interaction for the plasmonic and nanophotonic systems with potential applications in high density optical data storages.

While the diffraction of electromagnetic waves through an aperture in a metallic film is considered as a classical problem in electrodynamics, recent discoveries of extraordinary transmission (EOT) through subwavelength holes in thin metallic films at optical wavelengths [1,2] have challenged the famous results derived by Bethe in 1944 [3]. An enhanced transmission, of three orders of magnitude higher intensity than Bethe's theoretical prediction was observed, which triggered intense discussion in the last decades. In order to fully describe the microscopic picture of electromagnetic waves interacting with subwavelength structures, one must take into account both the intensity and phase information of the scattered fields. Many researchers have attempted to provide accurate explanations via theoretical studies [4], but few experimental studies are found in current literature on retrieving these information of the electromagnetic wave interacting with an isolated nanostructure in a metallic film.

Light emerging from an aperture on an opaque screen is a classical example of Huygens principle, which has been analyzed theoretically for centuries by researchers such as Grimaldi [5], Kirchhoff [6], Bethe [3], Bouwkamp [7], with excellent experimental agreements found in the microwave frequency region [8,9]. However, the classical theory of the light transmitted through a subwavelength aperture fails to explain the EOT phenomenon at optical wavelengths. The enhanced transmission is now attributed to resonant excitation of delocalized Bloch-state surface plasmon polariton waves (SPPs) in a one-dimensional periodic subwavelength slit case [10,11], as well as localized surface plasmon (LSP) modes of individual apertures in a two-dimensional case [10,12]. However, inconsistent results are observed when phase information in the subwavelength systems is studied [13-15].

In this letter, we present a near-field holographic system to retrieve both the intensity and phase information of the light diffracted from an isolated subwavelength structure in a thin metallic film. This study differs from existing literatures [16,17] whereby only the far-field intensities are measured. Our experimental observations do not fit into predictions from the classical theory of the light diffraction. In order to explain these experimental results, a horizontal electric dipole must be introduced in a hole surrounded by realistic metallic film. The presence of this dipole is not mentioned in the classical theory. This model can also predict that the excited dipoles in subwavelength dents and protrusions have opposite phases, which are confirmed by our experiments.

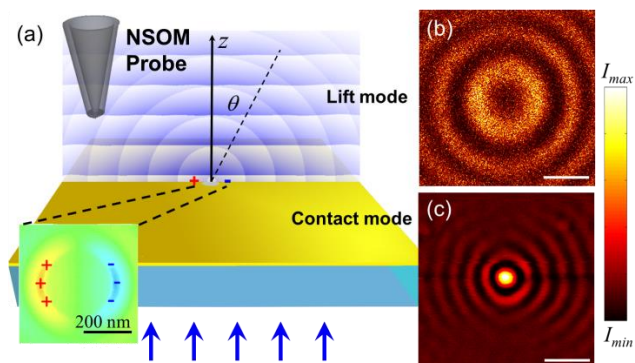

FIG. 1 (color online). (a) A schematic view of experimental configuration and holographic model. The inset is the simulated vertical field distribution right above the subwavelength hole. 300 nm diameter isolated hole in 100 nm thick Au thin film is illuminated by 482 nm light. (b) and (c) Optical intensity distribution for lift mode and contact mode, respectively. The scale bars are 1 μm.

In the system illustrated in FIG. 1(a), we use a near-field scanning optical microscopy (NSOM) to investigate optical field distribution near the isolated nanostructures in a thin Au film. A detailed sample preparation procedure can be found in the Supplementary Information [18]. Tapered multimode optical fibers coated by Al with 80-100 nm tip apertures, are used as scanning probes to collect the light. The apertured tips are used in our experiment to measure the optical intensity distribution around the hole, not only on the surface of the Au film, but also over different parallel planes within a few microns away from the surface. As shown in FIG. 1(a), the samples are illuminated through the substrate by a collimated laser beam of variable wavelength, which can be approximated as a plane wave illumination. In the experiment, we choose 482 nm and 647 nm laser lines from a Krypton laser to illuminate the samples. FIG. 1(b) and (c) show the pattern of optical intensity distribution measured on the sample of a 300 nm isolated hole in a 100 nm film. The two measurements are taken at the surface of the Au film (contact mode) and a height of 1 μm above the Au surface (lift mode), respectively. We can observe concentric rings in both the contact and lift mode images.

In order to explain the concentric-ring patterns recorded by the NSOM, we employ a holographic model, corresponding to the interference of a plane wave and a spherical wave (see FIG. 1(a)). The concept of interferometry has been applied in nanoparticle detection [19], single molecular imaging [20], etc., which studied far field signal, but we focus on the near-field interference patterns to explore the interaction of light with nanostructures. In our model, the plane wave $E_{plane} = A\exp(ikz)$ describes the field transmitted directly through the thin metallic film (although with significant attenuation). On the other hand, since our measurements are taken several wavelengths above the aperture, the radiation of the induced dipole in the aperture can be approximated by a (scalar) spherical wave $E_{spherical} = B\frac{\exp(ikr + i\varphi)}{r}$. Here, we introduce a phase shift $\varphi$ between the plane wave and spherical wave, and Eq. (1) captures the resulting intensity.

$$I(z,\phi) = (E_{plane} + E_{spherical}) \times (E^*_{plane} + E^*_{spherical})$$
$$= A^2 + \left(\frac{B\cos\theta}{z}\right)^2 + 2AB\frac{\cos\theta}{z}\cos[(kz - \frac{kz}{\cos\theta}) + \varphi], \quad (1)$$

Based on Eq. (1), we can deduce the complex polarizability of the dipole moment excited on the single subwavelength hole as:

$$\alpha = \frac{4\pi}{Z_0 ck^2} \frac{B}{(A/t)} \exp(i\varphi'), \quad (2)$$

where $Z_0$ is the wave impedance of light in the air (~377 Ω), $c$ is the light speed, $k$ is the wavevector, $t$ is a factor that describes attenuation of incoming light after transmitted through the film, and $\varphi'$ is phase difference between the incident wave and the excited dipole moment. $A$, $B$ and $\varphi'$ are unknowns that are determined from experimental measurements. Using the above holographic model, we then quantify the complex polarizability $\alpha$ of the isolated hole as a function of hole size, the thickness of the Au thin film, as well as the illumination wavelength through systematic experiments. The detailed method to retrieve the parameters is described in the Supplementary Information [18].

While the setup is similar to a recent report [21] on near field imaging of SPP through metallic holes, our aim here is to quantitatively study the induced horizontal electric dipole to elaborate the connection of the scattered fields to the illuminating field. This helps to reveal the microscopic process of light-structure interaction. In Bethe's original theory of light transmission through subwavelength aperture [3], all tangential electric field on the top surface of the metal should vanish, if the film can be regarded as an ideal conductor. In that case no interference pattern should be observed right on top of thin metallic film, since diffracted electric field is orthogonal to the transmitted field at incidence. However, we experimentally observe clear interference patterns under contact mode (shown in FIG. 1(c)), in agreement with Ref [21]. An additional horizontal electric dipole moment, which is not mentioned in the previous model, has to be introduced in order to match the diffracted electric field from the subwavelength hole in the metallic film. Our own full wave simulation result also confirmed the excitation of such in-plane dipole moment near the isolated hole (shown in the inset of FIG. 1(a)), in accord with the computational prediction by Ref. [22]. Moreover, we can observe concentric rings with a dark spot at the center in the lift mode image, which cannot be explained by the diffraction theory of a tiny hole.

By applying a non-linear fitting method on the experimental results (shown in FIG. S1), we retrieve phase shift $\varphi$s between the plane wave and the spherical wave which are around 100°. According to Eq. (1), the intensity at the center ($\theta = 0$) of the interference pattern is dominated by the phase shift $\varphi$. The dark spot observed at the center is caused by the retrieved dramatic phase change. We analytically calculate a $\Delta\varphi$, which describes the phase difference of two paths: i) the light passing through the thin metallic film by calculating the phase changes at both interfaces (Au-quartz and Au-air) and optical path through the thin film; ii) the light passing through the air layer of the same thickness. Varying the thickness of Au thin film from 0 to 120 nm, the $\Delta\varphi$s are within 10 degrees for 482 nm line. The detailed results are shown in the Supplementary Information [18]. This shows that the phase shift is caused by the horizontal electric dipole excitation. We calculate the corresponding phase shift $\varphi'$ from the retrieved $\varphi$ in the experiment. In addition, the total energy transmitted through the subwavelength hole towards the upper space is half of the dipole radiation. By calculating the dipole radiation power normalized by the incident energy, we can estimate the total transmission coefficient per unit area $T = \frac{4}{3}\frac{B^2}{r_0^2(A/t)^2}$, where $r_0$ is the radius of the hole in the metallic film.

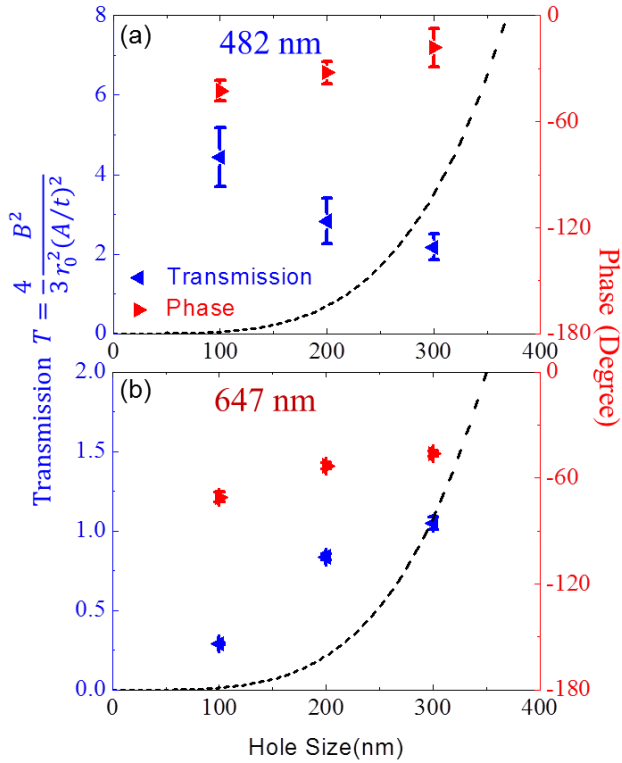

FIG. 2 (color online). The hole size dependent transmission coefficient per unit area (blue) and the corresponding phase shift (red) of the dipole moment illuminated by wavelength 482 nm (a) and 647 nm (b), respectively. The dashed curve represents the transmission coefficient per unit area from Bethe's theory.

FIG. 2(a) and (b) show the normalized transmission coefficient per unit area and the corresponding phase shift under different illumination wavelengths (482 nm and 647 nm). The hole sizes are varied from 100 nm to 300 nm. According to the calculation of cut-off frequency in cylindrical waveguide, there is no guided mode in such small subwavelength holes. We also plot the $(kr)^4$ dependence from Bethe's theory as dashed curve, which is valid for holes smaller than a quarter wavelength [23]. For the illumination by 482 nm light and a hole size of 100 nm, the transmission coefficient is two orders of magnitude larger than the prediction in Bethe's theory. And the transmission decreases when the hole size increases. However, for the 647 nm illumination, the transmission coefficient for the 100 nm hole is only several times larger than Bethe's theory, and this increases with the hole size. Since there is no guided mode in the cylindrical hole, the diffracted wave through the hole must be due to the SPPs exited at both interfaces and the induced in-plane dipole along the hole diameter. In our calculation, we use frequency-dependent permittivity data for Au from Palik [24]. At 482 nm and 647 nm, the permittivity of Au is -1.94+4.17i and -9.79+1.09i, respectively. We can see that it is close to the surface plasmon resonance at 482 nm, so that a strong horizontal dipole can be generated near the isolated subwavelength hole, which dramatically enhances the transmission through the hole. As the hole size increases, the strength of the dipole increases rather slowly, that leads to the decrease of the transmission per unit area. On the other hand, at 647 nm, the Au performs more like an ideal metal in Bethe's theory, so the trend of the transmission is similar to Bethe's prediction. The enhancement is still due to the SPPs and the induced dipole. But the strength of the dipole is much smaller compared to the 482 nm cases. In addition, the phase differences between the incident wave and the induced dipole are around $-\pi/6$ and $-\pi/3$ for 482 nm and 647 nm illumination, respectively, clearly distinguished from Bethe's and Kirchhoff's model. If the phase changes of the wave across the metal and quartz substrate, which are -34.5° and -61.2° by theoretical calculation for 482 nm and 647 nm, are taken into account, it is clearly seen that the induced dipole near the hole exactly follows the incident plane wave in the metallic film.

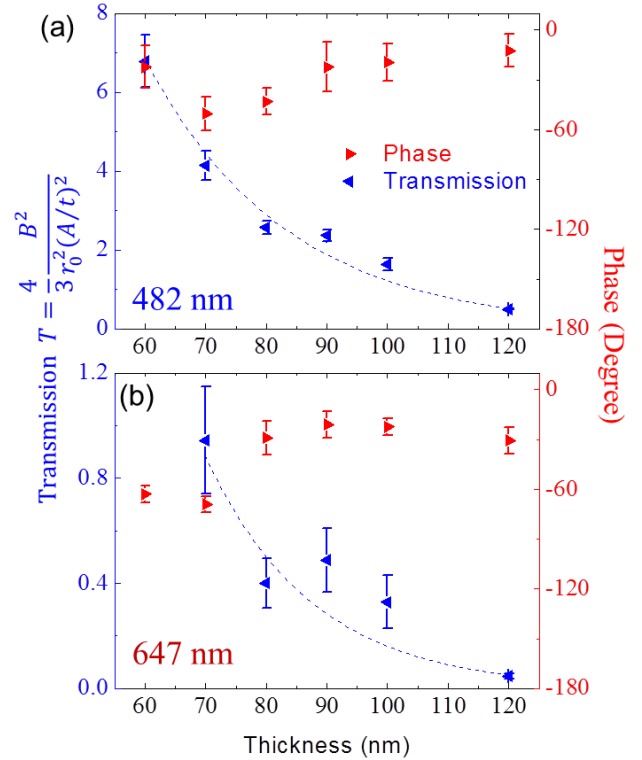

FIG. 3 (color online). The film thickness dependent transmission coefficient per unit area (blue) and the corresponding phase shift (red) of the dipole moment illuminated by wavelength 482 nm (a) and 647 nm (b), respectively. The dashed curve represents the exponential decay fitting for the transmission coefficient.

Furthermore, we study the thickness-dependent transmission and phase shift by fixing the size of isolated hole, and plot the retrieved transmission and phase shift in FIG 3(a) and (b) for 482 nm and 647 nm illumination, respectively. In the experiment, the thickness of the film with the isolated subwavelength hole (300 nm in diameter) varies from 60 nm to 120 nm. We can observe that the transmission decreases exponentially with the thickness of the Au film, with an exponential decay fit marked by dashed curves. The transmission coefficient decay length for 482 nm and 647 nm is 23.2 nm and 17.6 nm, respectively, in good agreement with theoretical calculations of 21.2 nm and 16.4 nm for the electric intensity decay length in Au. Similar to the phase shift in the hole size dependent cases, the phase of the excited dipole follows the incident plane wave. Therefore, we conclude that an in-plane electric dipole is exited at the incident interface by the incoming plane wave. The amplitude of the wave radiating from the dipole reduces across the metallic film, together with retarded phases.

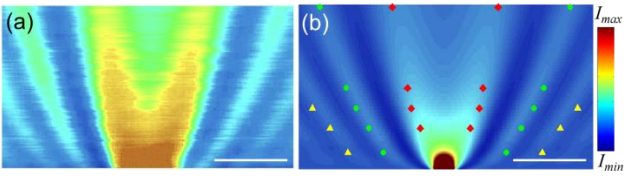

FIG. 4 (color online). (a) Measured cross section intensity distribution by vertical scanned NSOM in the sample of 300 nm isolated hole on 80 nm Au thin film illuminated by 482 nm laser light. (b) Simulated cross section intensity distribution by FDTD for the same configuration. The colored symbols are the interference peak positions captured from the intensity profile for the different heights. The scale bars are 1 μm.

We further explore the in-plane dipole radiation from near field to far field using the same NSOM system, and plot the cross section intensity distribution as a function of $z$ in FIG. 4(a). We choose a 300 nm isolated hole on an 80 nm Au thin film sample illuminated by the 482 nm laser line. The vertically scanned NSOM image clearly shows the radii change of the different order of the interference pattern. We also apply the full wave numerical simulation based on finite difference time domain (FDTD) method to simulate intensity distribution near the isolated hole in the Au metal thin film to confirm the experimental observation. The simulated intensity distribution of the cross section is plotted in FIG. 4(b). The color-coded symbols in FIG. 4(b) are the measured peak intensity positions superimposed on simulated results for various heights, showing good agreement between measurements and simulations.

Based on our experimental observation, we conclude that a horizontal electrical dipole moment near the hole is excited by an external electric field, and radiates. The phase of the diffracted wave is determined by the phase of the dipole moment. We can modify the phase by changing the structure of the subwavelength scatter. Considering an isolated dent and protrusion with the same size in the same thickness of Au thin film (see FIG. 5 insets), we can predict a charge accumulation near the subwavelength structure as pointed out in the figure. It is straightforward to observe that the directions of the excited dipole moments are opposite in the two cases. Based on the polarizability of the dipoles, we can predict a 180°phase difference in these two cases. This is confirmed by the intensity distribution on top of a dent and a protrusion in the same piece of 100 nm thick Au thin film (shown in FIG. 5). The two intensity distributions show an opposite relation, which means the direction of the dipole moments have been reversed. Applying the same data analysis method, we retrieve the phases for the dent and protrusion cases, which are 3.4° and 189.8°, respectively. The difference between these two phase is close to an expected 180°, which verifies our dipole moment prediction of the phase. Our result is similar to the famous Babinet's principle of complimentary fields in optics, in which the radiation patterns caused by the complementary structures display opposite phase. Considering the depth and thickness of the dent and protrusion (40 nm), which are much smaller than the wavelength, the phase shift observed is quite remarkable. Such dramatic phase shift due to the enhanced light-matter interaction by local resonance in subwavelength structures, might find exciting applications in high density optical data storage.

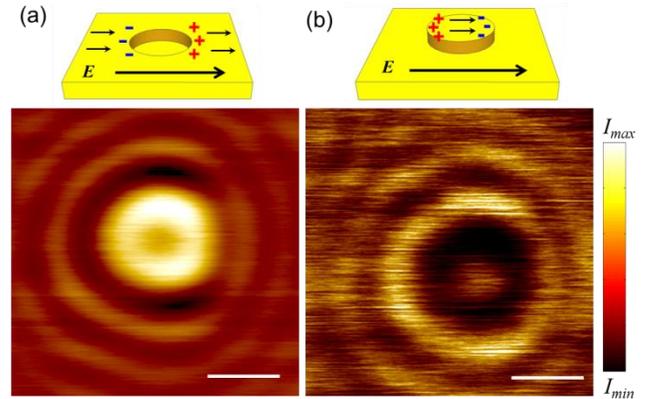

FIG. 5 (color online). (a) and (b) are the NSOM captured optical images of an isolated dent and protrusion on the Au thin film at the height of 1 μm, respectively. The scale bars are 1 μm. The insets are schematic view of the experimental configuration. The electric field direction and induced charge accumulation near the isolated dent and protrusion in the thin Au film are identified.

In summary, we have experimentally studied the electric field transmitted through an isolated subwavelength hole in metal thin film by analyzing the interference pattern captured by the NSOM setup. We observed an in-plane electric dipole moment excited by the external field which is not included in Bethe's model. Our holographic model enables us to retrieve both the transmission coefficient and phase shift information of the excited dipole near the hole. The transmission coefficient per unit area can be two orders larger than the previous theoretical prediction when strong surface plasmon resonance is exited. According to the experimental results, the transmission coefficient depends on both the hole size and the metallic film thickness. The phase of the exited dipole moment follows the illuminated electric field and can be tuned by local structure. Our description of the light in the isolated subwavelength hole is useful in understanding the essential mechanism of extraordinary transmission and local field distribution of subwavelength structure. The phenomenon shows the potential applications for phase and intensity control in subwavelength systems.

The authors thank K. H. Fung for useful discussion. We are grateful to N. Boechler, D. Jin and Z. J. Tan for a critical reading of the manuscript. We also acknowledge the financial support by the NSF (grant CMMI 0846771, CMMI-1120724) and AFOSR MURI (Award No. FA9550-12-1-0488).

*nicfang@mit.edu

# Supplementary Information: Near-field Holographic Retrieval of an Isolated Subwavelength Hole in a Thin Metallic Film

Jun Xu[1,2], Hyungjin Ma[3], and Nicholas X. Fang[1]
[1]Department of Mechanical Engineering, Massachusetts Institute of Technology, Cambridge, MA, 02139
[2]Department of Mechanical Science and Engineering, University of Illinois at Urbana-Champaign, Urbana, IL, 61801
[3]Department of Physics, University of Illinois at Urbana-Champaign, Urbana, IL, 61801

## Method for Retrieving Complex Polarizability

In FIG. S1, we show the intensity profiles retrieved from the NSOM measurement. With a sample of isolated 300 nm diameter hole on a 120 nm thick Au film, we measure the intensity distribution by the NSOM on different heights under the illumination from a 482 nm laser line. FIG. S1(a)-(d) insets show the NSOM optical images taken at the height of 500 nm, 750 nm, 1 μm, and 2 μm, respectively. We plot the intensity profiles (FIG. S1(a)-(d) in circles) by integrating the optical intensity at the same distance away from the center. We can observe that as the measurement height increases, the radii of rings become larger and larger, but the center part remains dark.

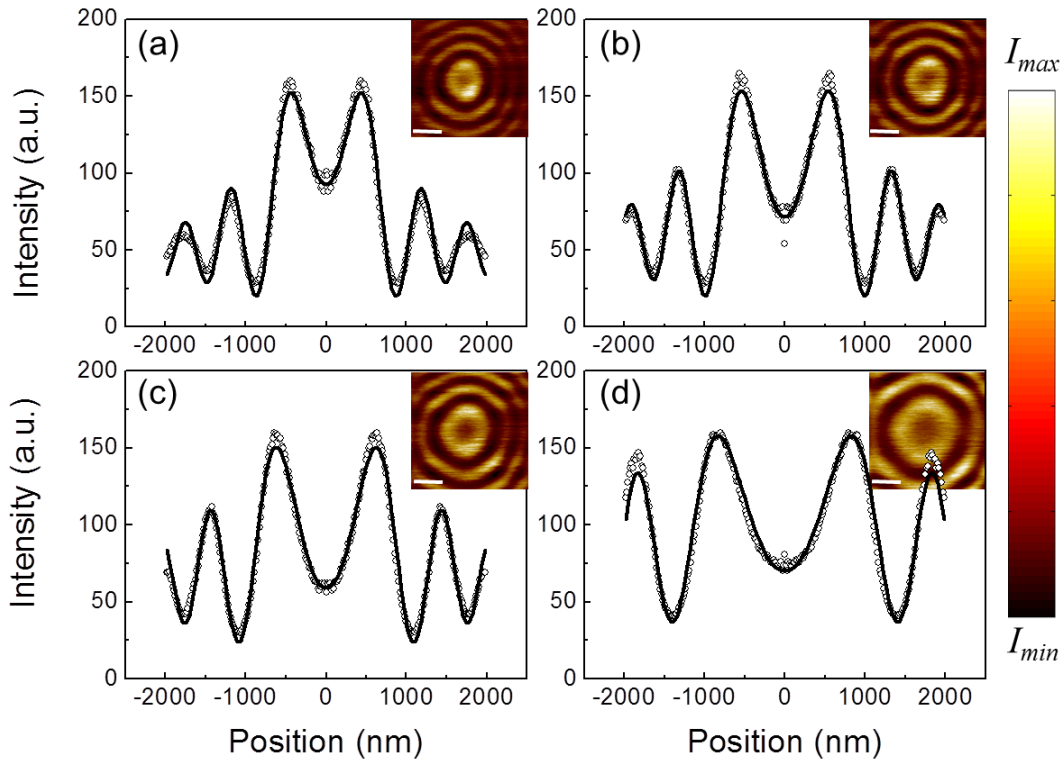

FIG. S1 The intensity distribution of 300 nm diameter isolated hole in 120 nm thick Au film measured by NSOM at the height of 500 nm, 750 nm, 1 μm, and 2 μm, respectively. The insets are the captured optical images. The intensity profiles obtained from the corresponding NSOM images are plotted in dots and nonlinear fitting curves are drawn in lines. The scale bars in the images are 1 μm.

According to Eq. (1), we apply a non-linear fitting process by treating $\varphi$, $A$, $B$ and $z$ as fitting parameters to fit the experimental intensity profiles, and draw the profiles in solid lines in the corresponding figures. It is clearly seen that the fitting curves match the experimental data very well. Therefore, our holographic model can describe this isolated subwavelength hole system accurately. The retrieved phase shift $\varphi$s are 89.9°, 101.9°, 109.8° and 110.0°, respectively. We should note that the $\varphi$ here is the phase difference between the transmitted plane wave and diffracted spherical wave as shown in Eq. (1). We need to consider the additional phase retardation of the incident wave passing through the thin metallic film to retrieve the phase difference $\varphi'$ that is between the incident wave and the excited dipole. In addition, we can retrieve the ratio between the amplitude of the plane wave $A$ and the spherical wave $B$, which is related to the energy transmission through the single subwavelength hole. By comparing the electric field radiation from a dipole (Eq. S1) and the spherical wave field distribution (Eq. S2), the complex polarizability ($\alpha$) of the horizontal dipole moment can be written as the function of the fitting parameters, as shown in Eq. (2).

$$E_{dipole} = Z_0 \frac{ck^2}{4\pi}(\alpha A_0)\frac{\exp(ikr)}{r}, \tag{S1}$$

$$E_{spherical} = B\frac{\exp(ikr - i\varphi)}{r}, \tag{S2}$$

Consequently, we can achieve the complex polarizability of the dipole excited on the subwavelength hole by substituting the retrieved $A$, $B$ and $\varphi'$. In Eq. S1, $A_0$ is the amplitude of the incident field. The amplitude of the transmitted plane wave $A$ equals $t*A_0$, where $t$ is the transmission coefficient. It can be analytically calculated based on the film thickness. To account for the variation of the actual height $z$ in the measurements, we also take height value as a fitting parameter in the non-linear fitting process. The retrieved height is around 10% higher than the input value, which is reasonable in the NSOM height control system.

Furthermore, the transmission coefficient per unit area of the light passing through the isolated subwavelength hole in the thin metallic film is calculated based on the dipole radiation power from the horizontal dipole moment :

$$P_{dipole} = \frac{c^2 Z_0 k^4}{12\pi}|\alpha A_0|^2 = \frac{4\pi B^2}{3Z_0}, \tag{S3}$$

Half of the radiation power, which radiates to the upper space, contributes to the energy transmission through the subwavelength hole, while the incident power is $P_{inc} = A_0^2/2Z_0$. So the total transmission coefficient per unit area can be written as:

$$T = \frac{4}{3}\frac{B^2}{r_0^2(A/t)^2}, \tag{S4}$$

where $r_0$ is the radius of the hole in the metallic film.

## Analytical Calculation of Light Passing through a Thin Metallic Film

We analytically calculate the electromagnetic wave passing through a thin Au film on top of the quartz substrate by using the frequency dependent permittivity of Au. The real part of the electric field provides the transmission coefficient $t$, and the imaginary part provides the phase retardation information. Here we plot the calculated phase difference between the EM waves passing through the Au thin film and air with the same thickness for different wavelength. It is found that under 482 nm illumination, the phase difference is quite small (< 15 degree) for the different thicknesses, but the experimental observation is around 100 degree. Therefore, an additional phase retardation mechanism must be involved in the light diffracted from an isolated subwavelength hole in the thin metallic film.

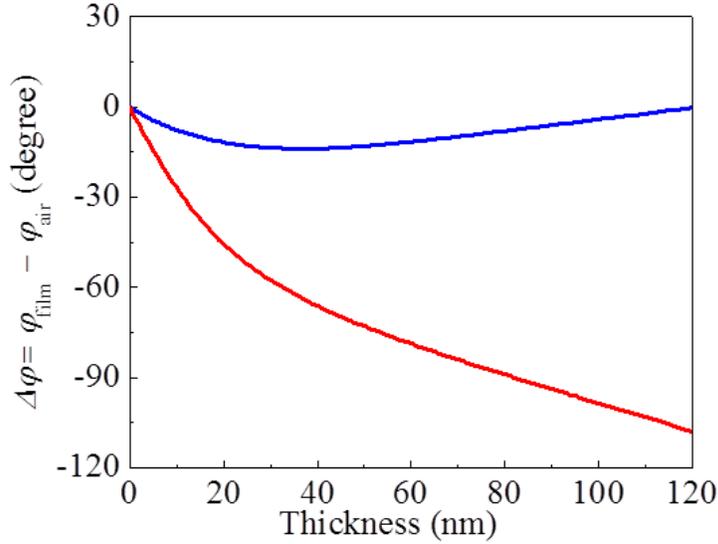

FIG. S2 Analytically calculated phase difference between light passing through the varied thickness of Au thin film and corresponding thickness of air for 482 nm (blue) and 647 nm (red) illumination.

## Full-wave Simulation

We use commercial software COMSOL® Multiphysics to numerically calculate the power radiation from an isolated subwavelength hole in the thin metallic film, and compare to our experimental results. A 3D simulation domain (3×3×3 μm) is constructed in the software. We choose 80 nm thick Au film on quartz substrate, and vary the diameter of the isolated hole from 100 nm to 300 nm to simulate the experimental cases plotted in FIG. 2. A plane wave illuminates the sample from the substrate side, and a 2D monitor is placed at air side to collect the radiation flux. Perfect matched layers (PMLs) are placed around the whole simulation domain to avoid reflection from the boundaries. We simulate both 482 nm and 647 nm illumination cases. The total radiation power integrated over the monitor contains two parts: the directed transmission through the thin metallic film and the power radiated from the subwavelength hole. In the simulation, we can easily subtract the first part, and retrieve the total radiation power only from the hole. We normalize the radiation power by the incident energy in the hole to achieve the total transmission coefficient per unit area, and plot them in FIG. S3 (green symbols). We can observe that the simulation results quantitatively agree with the experimental results. In the 482 nm case, the transmission coefficient drops as the hole size increases, while in the 647 nm case, it reverses. The discrepancy might be due to the experimental uncertainties including the thickness of Au film, the hole size, and rounded edge of the hole drilled by focused ion beam, etc.

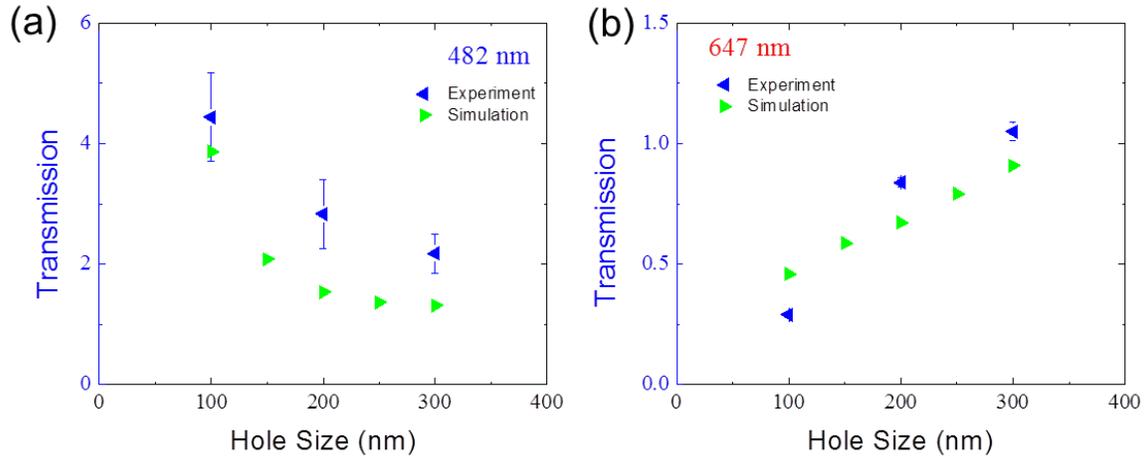

FIG. S3 The comparison of experimental (blue) and simulation (green) results of the hole size dependent transmission coefficient per area. The samples are illuminated by wavelength 482 nm (a) and 647 nm (b), respectively.

## Sample Fabrication Methods

The Au films are deposited onto 300 μm fused quartz substrates by electron beam evaporation. The thickness values of the Au films varied from 60 nm to 120 nm. We then drill isolated holes with various diameters (100 nm to 300 nm) by focused ion beam milling. The typical SEM image of the isolated hole sample is shown in FIG. S4(a). For the protrusion sample, after depositing 100 nm Au film on the substrate, we pattern the PMMA template by using by electron beam lithography method. Followed by the second-time Au deposition and lift-off process, we achieve the protrusion with 300 nm diameter and 40 nm height in the Au film (FIG. S4(b)). Then we drill an isolated dent with 300 nm diameter and 40 nm depth by focused ion beam milling on the same piece of sample to retain the film thickness.

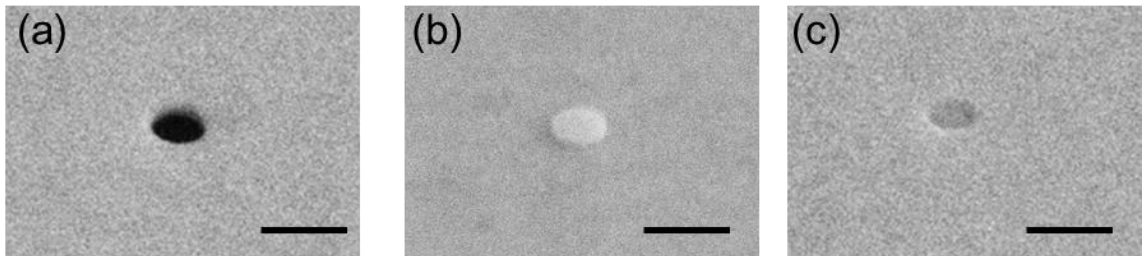

FIG. S4 SEM images of isolated hole (a), protrusion (b) and dent (c) on the thin Au film. The scale bars are 500 nm.